\documentclass[twocolumn,trackchanges]{aastex63}
	\usepackage{color}
	\usepackage{multirow}
	\usepackage{amsmath}
	\usepackage{lineno}

	\received{Month date, 2020}
	\revised{Month date, 2020}
	\accepted{Month date, 2020}
	\shorttitle{Sample article}
	\shortauthors{Hao Cheng et al.}
	
\begin{document}

\title{Photo-nuclear reaction rates of $^{157,159}$Ho and $^{163,165}$Tm and their impact in the $\gamma$--process}

\correspondingauthor{Bao-Hua Sun}
\email{bhsun@buaa.edu.cn}
			
\correspondingauthor{Li-Hua Zhu}
\email{zhulh@buaa.edu.cn}
			
\correspondingauthor{Yudong Luo}
\email{yudong.luo@pku.edu.cn}

\author{Hao Cheng}
\affiliation{China Institute of Atomic Energy, Beijing 102413,  P. R. China\\}

\author{Bao-Hua Sun}
\affiliation{School of Physics, and International Research Center for Big-Bang Cosmology and Element Genesis, Beihang University, Beijing 100191, P. R. China}

\author{Li-Hua Zhu}
\affiliation{School of Physics, and International Research Center for Big-Bang Cosmology and Element Genesis, Beihang University, Beijing 100191, P. R. China}
			
\author{Motohiko Kusakabe}
\affiliation{School of Physics, and International Research Center for Big-Bang Cosmology and Element Genesis, Beihang University, Beijing 100191, P. R. China}

\author{Yudong Luo}
\affiliation{School of Physics, Peking University, Beijing 100871, China}
\affiliation{Kavli Institute for Astronomy and Astrophysics, Peking University, Beijing 100871, China}

\author{Toshitaka Kajino}
\affiliation{School of Physics, and International Research Center for Big-Bang Cosmology and Element Genesis, Beihang University, Beijing 100191, P. R. China}
\affiliation{National Astronomical Observatory of Japan, 2-21-1 Osawa, Mitaka, Tokyo 181-8588, Japan}
\affiliation{Graduate School of Science, The University of Tokyo, 7-3-1 Hongo, Bunkyo-ku, Tokyo 113-0033, Japan}

\author{Chang-Jian Wang}
\affiliation{School of Physics, and International Research Center for Big-Bang Cosmology and Element Genesis, Beihang University, Beijing 100191, P. R. China}
			
\author{Xing-Qun Yao}
\affiliation{School of Physics, and International Research Center for Big-Bang Cosmology and Element Genesis, Beihang University, Beijing 100191, P. R. China}
			
\author{Chuang-Ye He}
\affiliation{China Institute of Atomic Energy, Beijing 102413,  P. R. China\\}
			
\author{Fu-Long Liu}
\affiliation{China Institute of Atomic Energy, Beijing 102413,  P. R. China\\}
			
\author{Bing Guo}
\affiliation{China Institute of Atomic Energy, Beijing 102413,  P. R. China\\}


			
			
			
\begin{abstract}
				
Reliable photo-nuclear reaction rates at the stellar conditions are essential to understand the origin of the heavy stable neutron--deficient isotopes between$^{74}$Se and $^{196}$Hg—$p$-nuclei, however, many reaction rates of relevance still have to rely on the Hauser-Feshbach model due to rare experimental progress. One such case is in the mass range of 160 for Dy, Er, Ho and Tm isotopes.
In this work we attempt to constrain the Hauser-Feshbach model in the TALYS package by reproducing the available experimental data of $^{160}$Dy($p,\gamma$)$^{161}$Ho and $^{162}$Er($p,\gamma$)$^{163}$Tm  in the $A\sim 160$ mass region, and examine the effects of level density, gamma strength function and the optical model potential. The constrained model then allows us to calculate the reaction rates of $^{157, 159}$Ho($\gamma$, $p$) and $^{163,165}$Tm($\gamma$, $p$) for the $\gamma$-process nucleosynthesis in carbon-deflagration SNe Ia model. 
Our recommended rates differ from the JINA REACLIB by more than 1 order of magnitude in the temperature range of 2--3 GK. This 
results in the changes of final abundance of $p$-nuclei in the $A\sim 160$ mass range by -5.5--3\% from those with JINA, which means that the ($\gamma$, $p$) reactions uncertainty is not predominant for the synthesis of these nuclei.
\end{abstract}

			
\section{Introduction} \label{intro}
35 neutron--deficient stable nuclei between $^{74}$Se and $^{196}$Hg, called  $p$-nuclei, cannot be synthesized via the $s$ and $r$ neutron-capture processes. One of the most favored production mechanisms for those nuclei is photodisintegration, the so-called $\gamma$--process~\citep{Woosley1978The, 1990A&A...227..271R, 1991ApJ...373L...5H,1995A&A...298..517R,2003ApJ...585..418F,2023arXiv230611409R}.
Core--Collapse Supernovae (CCSNe) and the thermonuclear supernovae (SNe Ia) are two main astrophysical sites where one believes that the $\gamma$--process could occur~\citep{2011ApJ...739...93T,2014ApJ...795..141T,2018ApJ...854...18T}.

The $\gamma$--process in CCSNe can produce $p$--nuclei in various mass ranges, and a recent study indicates that different progenitor models of CCSNe may provide different isotopic ratios of $p$-nuclei compared with the solar abundances~\citep{2023arXiv230611409R}. Most $p$--nuclei in the mass ranges 124 $\leq$ A $\leq$ 150 and 168 $\leq$ A $\leq$ 200 can be produced in CCSNe, while there are severe deficiencies in the A $\leq$ 124 and  A$\sim$160 mass regions, where the solar abundances are severely under-produced~\citep{2001Nucleosynthesis,Sensitivity2006,2008ApJ...685.1089H,Rauscher2013Constraining,2013ApJ...762...31P}. 			
The $\gamma$--process could also occur in the explosions of carbon-oxygen white dwarf (WD) triggered by deflagration or detonation after accreting mass from a companion star \citep{1991ApJ...373L...5H, 2005NuPhA.758..459K,arn06,Kusakabe2010Production,2011ApJ...739...93T,2014ApJ...795..141T}. 
The extensive $s$-process in the progenitor asymptotic giant branch (AGB) star ~\citep{2020MNRAS.497.4981B} and the helium shell ﬂashes during the mass accretion could produce seeds nuclei for $\gamma$--process. However, even for a presumption of a factor of  $10^3-10^4$ overproduction of $s$-process seeds nuclei, the solar abundance pattern of $p$-nuclei could not be reproduced completely~\citep{2005NuPhA.758..459K,Kusakabe2010Production,Rauscher2013Constraining}. The abundances of nuclei in the mass range $A\sim160$ (i.e., $^{152}$Gd, $^{156,158}$Dy and $^{162,164}$Er) are particularly noteworthy since they cannot be accounted for within the $\gamma$-process nucleosynthesis models in both scenarios we mentioned above ~\citep{Rauscher2013Constraining}. 

Besides the two main scenarios, other astrophysical sites have also been discussed to potentially produce $p$-isotopes. For example, the isotropic fractions of $^{92, 94}$Mo and $^{96, 98}$Ru are 1--2 orders of magnitude larger than the other $p$--nuclei and their main production process has been suggested to $\nu p$--process~\citep{2006PhRvL..96n2502F,2011ApJ...729...46W,2022ApJ...924...29S}. A newly proposed $\nu r$--process could be responsible for the production of $p$-nuclei \citep{2024PhRvL.132s2701X}. Recent studies also suggest that $^{152}$Gd and $^{164}$Er are mainly reproduced by the $s$-process~\citep{2014ApJ...787...10B,2011ApJ...739...93T}.

The $p$-nuclei production is sensitive to their seed $s$- and (or) $r$-process nuclei, whose abundances strongly depend on the stellar evolution before the supernova explosion. Therefore, the precise prediction of the $p$-nuclei abundances requires many important factors among the uncertainties of both astrophysical model and the nuclear input parameters. However, they have not yet been clearly determined.
Modeling the synthesis of the $p$--nuclei and calculating their abundances~\citep{2022ApJ...941...56S} in the $\gamma$--process require an extended reaction network calculation involving about 10 000 reactions on around 2000 stable and unstable nuclei~\citep{Arnould2003The}.
The availability and uncertainties in nuclear data have a significant impact on the reaction rates and on the final abundance distribution of elements.
Despite the concerted experimental efforts~\citep{2020PhRvC.102d5811S,2020PhLB..80735575F,2021PhRvL.127k2701L,2021ApJ...915...78C,2022EPJWC.26011030V,2023EPJWC.27911004W,2023PhRvC.107c5803W,2023PhRvC.107c5808T}, most of the nuclei involved are currently unavailable in the laboratory  as they often involve unstable nuclei or extremely low isotopic fractions.  
Thus, reaction rates are frequently predicted relying on statistical Hauser-Feshbach (HF) theory models~\citep{Hauser1952The,1997PhRvC..56.1613R}.
Several popular software packages, such as TALYS ~\citep{2023EPJA...59..131K}, including various modifications of phenomenological and microscopic nuclear structure and reaction models, are extensively used nowadays for nuclear reaction calculations of different projectiles~\citep{2003PhRvC..67a5807U,2011PhRvC..84a5802D,MeiBo2015,2006PhRvC..74b5805G,2014PhRvC..90c5806N,2009PhRvC..80c5804G,2020PhLB..80535431W}.
However, the physical inputs inherent to the HF theory calculations may not have undergone well constrained verification. 
Ensuring the provision of reliable reaction rates necessitates conducting a validity test of the statistical HF model in the mass region of interest.

For the case of the $\gamma$-process, the dominant reactions are photodisintegration reactions (($\gamma$, $n$), ($\gamma$, $p$) and ($\gamma$, $\alpha$)). However, despite the challenges in obtaining quasi-monoenergetic $\gamma$-rays, in most cases cross sections of ($p, \gamma$) and ($\alpha, \gamma$) reactions have been measured  to deduce their inverse, ($\gamma, p$) and ($\gamma, \alpha$), which are related through the detailed balance theorem. 
For  experimental studies related to $p$-nuclei in the  A$\sim$160 mass region, very few proton capture measurements have been performed except for $^{162}$Er~\citep{2017as} and $^{160}$Dy~\citep{2021ApJ...915...78C}. 
Experimental data of $^{156}$Dy, $^{158}$Dy, and $^{164}$Er in this mass region are still unavailable due to the low cross section (0.3 nb $\sim$ 0.4 mb) at the energies of Gamow window and the low isotopic fractions (0.06\% $\sim$ 1.60\%). 
As an alternative, it's possible to utilize the known experimental data within a similar mass range to constrain the HF theoretical reaction models. 
The available cross section data  in the $A$$\sim$160 mass region are $^{160}$Dy($p$, $\gamma$) and $^{162}$Er($p$, $\gamma$), obtained at proton laboratory energies ranging from 3.5 to 9.0 MeV, covering  the majority of the Gamow window for each reaction.
Subsequently, a short-range extrapolation is conducted to directly predict the  photo-proton reaction rates of $^{157,159}$Ho and $^{165,163}$Tm, and to study their impact on the $\gamma$-process nucleosynthesis. 

The strong constraints on reaction models help to improve and guarantee the reliable results of the $\gamma$-process calculations. In the present paper, the experimental data for proton capture reactions on $^{160}$Dy and $^{162}$Er are compared with the TALYS-2.0 calculations obtained with various nuclear level density (NLD), optical model potential (OMP), gamma strength function (GSF), compound nucleus (CPN), and pre-equilibrium (pre-E). In Sec.~\ref{SC}, 
we present a comprehensive illustration of constraining the HF theory models using the experimental data, and identify the most influential nuclear models.
We also compare the extracted reaction rates of  $^{157,159}$Ho($\gamma, p$)$^{156,158}$Dy and $^{163,165}$Tm($\gamma, p$)$^{162,164}$Er with those from the JINA reaction library~\citep{2010ApJS..189..240C}.
In Sec.~\ref{SNyields}, we apply the new reaction rates to the nucleosynthesis in low-temperature layer of a carbon-deflagration model for SNe Ia, which is not included in the previous study \cite{Kusakabe2010Production}. We compare the $p$-nuclei final abundances between the derived rates and JINA REACLIB-V2.0 rates and discuss the overproduction factors of $p$-nuclei. Finally, we present the conclusions in Sec.~\ref{conclusion}.

			
\section{Hauser-Feshbach Calculations}\label{SC}
\subsection{Theoretical Model Framework}
Necessary conditions for applying the statistical HF model, such as high excitation energies involving many overlapping energy states, are readily satisfied for many nuclear reactions in the $\gamma$-process.  
All existing $\gamma$-process calculations actually rely on a statistical HF model.
For a reaction $A^{i}(a,b)B^{f}$,  where the target nucleus $A$ in initial level $i$ captures an incident $a$, forming a compound nucleus which subsequently decays to final level $f$ of residual $B$ with emission of particle $b$, with a center of mass energy $E_{Aa}$ and reduced mass $\mu_{Aa}$ for the entrance channel, the cross section $\sigma^{i,f}$ is given by:
\begin{equation}
	\begin{split}	
	\sigma^{i,f}(E_{Aa})=\frac{\pi\hbar^2/(2\mu_{Aa}E_{Aa})}{(2J^i_A+1)(2J_a+1)}\sum_{J,\Pi}(2J+1) \\ \times\frac{T^i_a(E,J,\Pi,E^i_A,J^i_A,\Pi^i_A)T^f_b(E,J,\Pi,E^f_B,J^f_B,\Pi^f_B)}{T_{tot}(E,J,\Pi)},
		\label{eq1}
	\end{split}
\end{equation}
where $J$ is the spin, $E$ is the corresponding excitation energy, and $\Pi$ represents the parity of excited states of residual $B$.
$T^i_a$ represents the transmission coefficient for formation of compound nucleus from entrance state $i$ of $a$, and $T^f_b$ is the transmission coefficient for evaporation of a particle $b$ to the final state $f$ of residual $B$ from compound nucleus.
The total transmission coefficient $T_{tot}=\sum_{f,b}T^f_b$ describes the transmission into all possible bound and unbound states $f$ in all energetically accessible exit channels $b$ (including the entrance channel $A$).
And $T^f_b$ can be replaced by transmission coefficient into exit channel $b$:
\begin{equation}
	\begin{split}	
	T_b(E,J,\Pi)=\sum_{f=0}^{f_B}T_b^f(E,J,\Pi,E^f_B,J^f_B\Pi^f_B)\\
	+\int_{E_B^{f_B}}^{E_B^{max}}\sum_{J_B.\Pi_B}T_b(E,J,\Pi,E_B,J_B,\Pi_B)\\
	\times	\rho(E_B,J_B,\Pi_B)dE_B,
		\label{eq2}
	\end{split}	
\end{equation}
where $E_B^{f_B}$ is the energy of the highest experimentally known bound excited state $f_B$. $\rho(E_B,J_B,\pi_B)$ is the NLD which describes the number of states with spin $J_B$ and parity $\pi_B$ at excitation energy $E_B$ .
			
Thus, the important quantities for calculating a cross section from Eqs.(\ref{eq1}) and (\ref{eq2}) are the transmission coefficients of particles and $\gamma$-rays, and the NLD. The particle transmission coefficients are calculated using an OMP, and the $\gamma$-transmission coefficients are governed by GSF parameters.
TALYS provides a variety of options for describing these inputs and is used hereafter to conduct the sensitivity studies.
			
\begin{figure}
	\centering
	\includegraphics[width=0.52\textwidth,clip=true,trim=0cm 0cm 0cm 0cm]{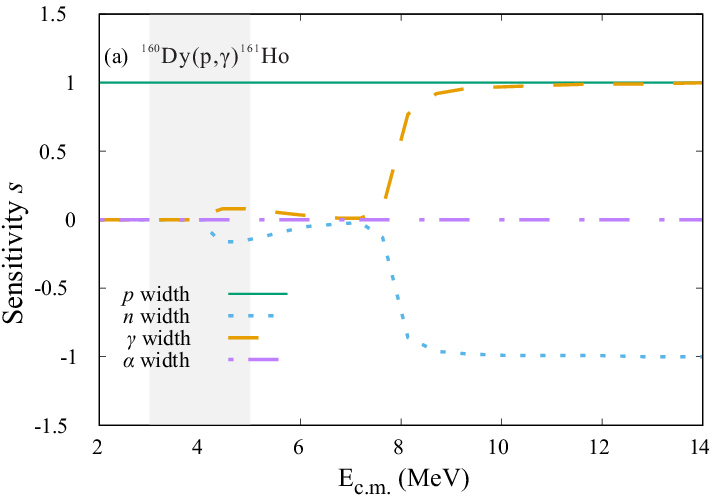}
	\includegraphics[width=0.52\textwidth,clip=true,trim=0cm 0cm 0cm 0cm]{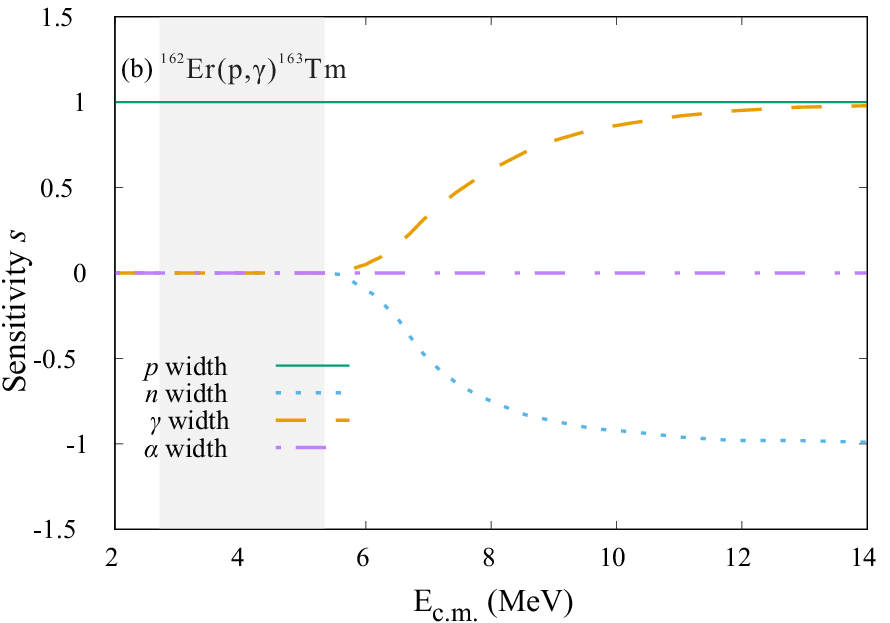}
	\caption{(Color online) Sensitivity $s$ of the $^{160}$Dy($p,\gamma$)$^{161}$Ho (a) and $^{162}$Er($p,\gamma$)$^{163}$Tm (b) corss sections, when the $p$, $n$,  $\alpha$, and $\gamma$ widths are varied by a factor of two, as a function of center-of-mass energies. The data are from ~\cite{2012ApJS..201...26R}.  The shaded area denotes the Gamow window. Within the Gamow windows, the cross sections are predominantly sensitive to the $p$ width.}
		\label{fig1}
\end{figure}

The sensitivity of the $^{160}$Dy($p$, $\gamma$) and $^{162}$Er($p$, $\gamma$) cross sections to charged-particle widths, neutron widths, and $\gamma$ widths varies with respect to the center-of-mass energy,  as depicted in Fig.\ref{fig1}. The sensitivity $s$ denotes the relative change of the cross section when one width is varied~\citep{2012ApJS..201...26R}:
\begin{equation}
	s=\frac{\frac{\sigma^{'}}{\sigma}-1}{f-1}
\end{equation}
In this context,  $\frac{\sigma^{'}}{\sigma}$ represents the relative variation of the laboratory cross section and $f=\frac{\Gamma^{'}}{\Gamma}$ signifies the factor that adjusts the respective width. The neutron threshold for $^{160}$Dy and $^{162}$Er are 4.1 and 5.6 MeV, respectively. The nuclear-physics input parameters include NLD and GSF which determine the $\gamma$ width. Moreover, the OMP are needed to describe the particle widths for protons and neutrons. $^{160}$Dy($p$, $\gamma$) and $^{162}$Er($p$, $\gamma$) cross sections are dominantly sensitive to the proton ($p$) width within the Gamow window. Above the neutron threshold, the effect of gamma ($\gamma$) and neutron ($n$) width increases with increasing energies.

\subsection{Constraining the Hauser-Feshbach Model}\label{Modelcalculations}
			
For the key ingredients  (NLD, OMP, GSF, etc) in Hauser-Feshbach calculations, TALYS offers a variety of options for the description of these inputs. However, these options presented in TALYS are not a systematic variation of nuclear properties and therefore cannot be used as measure of uncertainty in the cross section prediction.
It is beneficial to develop a sub-model combination for the mass region that can be applied to nearby nucleus where there is still limited or no available experimental data.
Multiple (sub-)models are required to calculate crucial input variables such as NLD, GSF, OMP, nuclear masses, pre-equilibrium, and the compound process to perform accurate nuclear reaction calculations.
We aim to find the optimal physical inputs for TALYS, which can accurately describe the known experimental data in $A\sim$ 160 mass region. 
			
\begin{figure}
	\centering
	\includegraphics[width=0.5\textwidth,clip=true,trim=0cm 0cm 0cm 0cm]{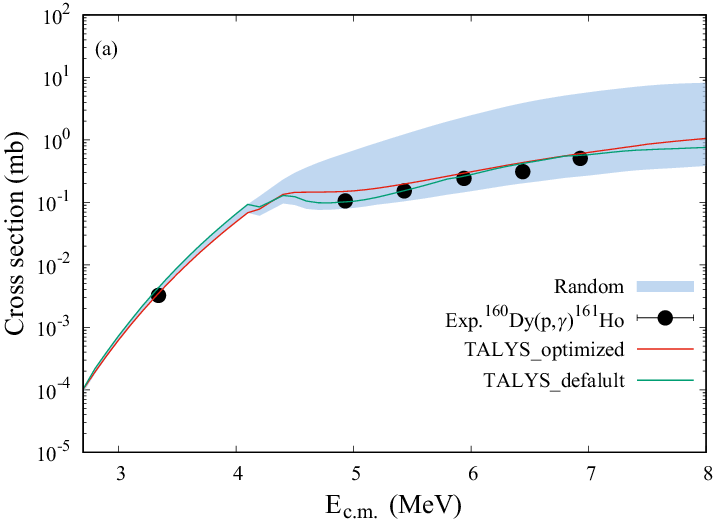}
	\includegraphics[width=0.5\textwidth,clip=true,trim=0cm 0cm 0cm 0cm]{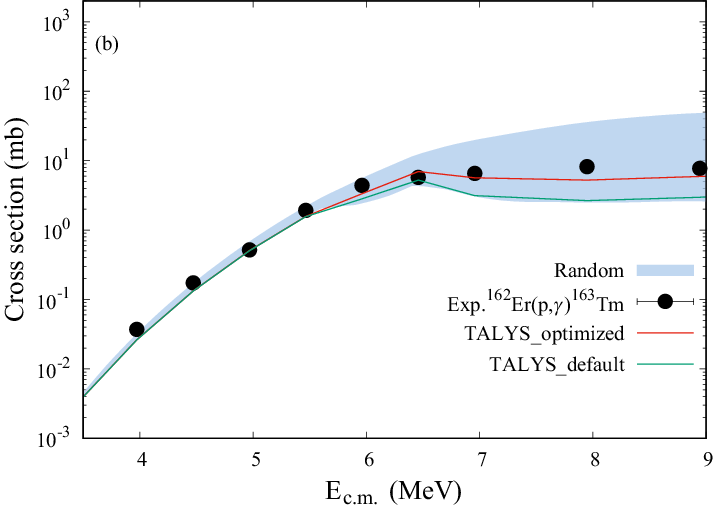}
	\caption{(Color online) TALYS calculation results compared with experimental data (black dots), for (a) $^{160}$Dy($p,\gamma$)$^{161}$Ho using default parameters (green line),  one set of sampling models (color filled areas), and the optimized one (red line) best reproducing experimental data both of $^{160}$Dy($p,\gamma$)$^{161}$Ho and $^{162}$Er($p,\gamma$)$^{163}$Tm reactions; (b) $^{162}$Er($p,\gamma$)$^{163}$Tm using default parameters (green line), sampling models (blue filled area) and the optimized one (red line). }
		\label{fig2}
\end{figure}

For the available cross section in the mass region around $A=160$,
we found that TALYS calculations with default inputs for $^{160}$Dy($p,\gamma$)$^{161}$Ho  are in good agreement with the experimental data at 3--9 MeV~\citep{2021ApJ...915...78C}, as shown in Fig.~\ref{fig2}(a). The TALYS result with the default parameters for the $^{162}$Er($p,\gamma$)$^{163}$Tm~\citep{2017as} reaction is a factor of two smaller than the experimental data at 6.5--9 MeV, as shown in  Fig.~\ref{fig2}(b). The experimental data and TALYS calculations with default inputs are shown in Table~\ref{tab:2}.
These calculations are expected to be a good starting point in extrapolations to the experimentally unknown mass region.
In order to identify the model combinations that can reproduce these experimental data, different physical models implemented in the TALYS 2.0 code are sampled and varied to generate random input files with unique model combinations.
			
 We perform calculations in TALYS by varying 52 key physical inputs, as described by~\cite{2019NDS...155....1K}. In total, 10 000 samples are calculated. The results create a large spread in distributions as shown in Fig.~\ref{fig2}. In calculations, the experimental mass values are used.  The results are well converged at low energies below neutron threshold, but deviates as the energy increases. 
			
The quality of the description of  experimental data with different model combinations is estimated using the reduced chi-square calculation:
\begin{equation}
	R^2=\frac{1}{N-1}\sum_{i=1}^{N} \frac{(\sigma^{exp}-\sigma^{cal})^2}{(\Delta\sigma^{exp})^2},
\end{equation}
where $\sigma^{exp}$ and  $\sigma^{cal}$  are the experimental and calculated cross sections, and $\Delta\sigma^{exp}$ is the uncertainty of experimental data. $N$ is the number of experimental data points.
The least $R^2$ is 9.3.
The reaction cross-section is most sensitive to the variations of the OMP, NLD, and GSF. Altering the OMP, NLD, and GSF result in changes in chi-square values by factors of up to 4, 17 and 11, respectively (see Table~\ref{tab:1} for the relevant inputs).

\subsection{Constraining the models}
			
For the NLD, the  Back-shifted Fermi gas model is used. The internal parameters constrained are the asymptotic level-density parameter $\widetilde{a}$, the level-density damping parameter $\gamma$ of the three nuclei in the proton-entrance, neutron-exit, and $\gamma$-exit channels of each reaction. For example, in the proton induced reaction on $^{160}$Dy, the final products, $^{160}$Dy, $^{160}$Ho and $^{161}$Ho, are considered.  
			
For the GSF, $E$1 transitions are calculated using the microscopic Hartree-Fock BCS tables. Regarding the OMP, the Koning and Delaroche OMP is used, which is suitable  for protons and neutrons with incident energies from 1 keV up to 200 MeV, for (near-)spherical nuclides in the mass range 24 $\leq$ A $\leq$ 209~\citep{2003NuPhA.713..231K}. The parameters within these models are not varied and kept as the TALYS default. Below the neutron threshold, all the parameters are well constrained by the experimental data and tend to be similar as shown in Fig.~\ref{fig2}. This also verifies the OMP can well describe the current reactions.
			
We compare the cross sections to TALYS predictions using different asymptotic $\widetilde{a}$ and damping gamma, while keeping all other ingredients to the calculations fixed. For $^{160}$Dy, $^{160}$Ho, $^{161}$Ho, the constrained $\widetilde{a}$ values are 17.14, 17.14, 17.23, respectively. All of the damping parameters are 0.072. The cross sections are mostly sensitive to the damping parameter $\gamma$ for Back-shifted Fermi gas level density model. The cross sections of $^{160}$Dy($p,\gamma$)$^{161}$Ho and $^{162}$Er($p,\gamma$)$^{163}$Tm calculated by the constrained model and parameters (“TALYS\_optimized") are shown in Fig~\ref{fig2}. TALYS\_optimized can be employed to evaluate the cross sections of the $A$$\sim$160 mass region.

\subsection{Evaluation of the Stellar Rate}\label{SR}

\begin{figure}[htbp!]
	\centering	\includegraphics[width=0.5\textwidth,clip=true,trim=0cm 0cm 0cm 0cm]{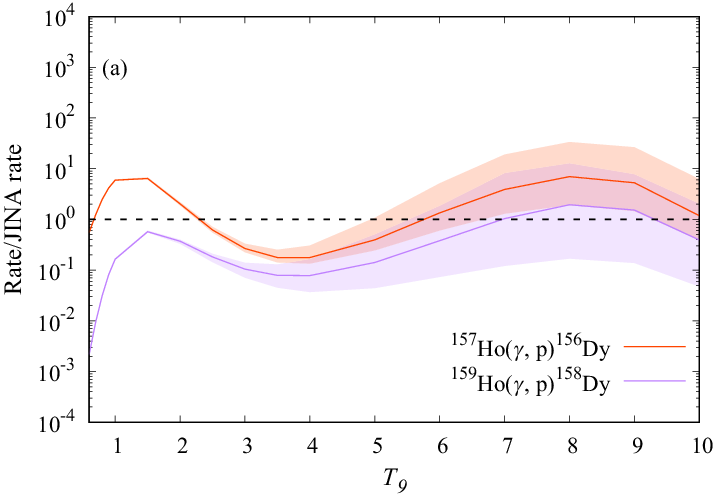}	\includegraphics[width=0.5\textwidth,clip=true,trim=0cm 0cm 0cm 0cm]{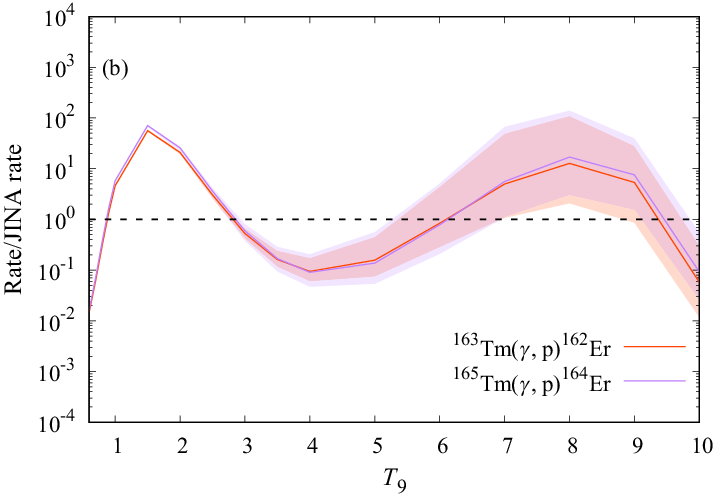}
	\caption{(Color online) (a) Ratios of $^{157}$Ho($\gamma, p$)$^{156}$Dy and $^{159}$Ho($\gamma, p$)$^{158}$Dy stellar rates deduced from the “TALYS\_optimized" and JINA (solid lines). The associated uncertainties are shown as the coloured bands. (b) The same as upper panel (a) but for the  $^{163}$Tm($\gamma, p$)$^{162}$Er and $^{165}$Tm($\gamma, p$)$^{164}$Er reactions. }
	\label{fig3}
	\end{figure}
			
For reaction rates calculations, the ground state and the excited states of the target nucleus is assumed to be in thermodynamic equilibrium within the hot astrophysical plasmas. The various nuclear levels with spins and excitation energies are assumed to obey a Maxwell-Boltzmann distribution.
The “TALYS\_optimized” calculation is found to well describe the experimental data and is used to determine the reaction rates of  $^{157}$Ho($\gamma, p$)$^{156}$Dy, $^{159}$Ho($\gamma, p$)$^{158}$Dy, $^{163}$Tm($\gamma, p$)$^{162}$Er and $^{165}$Tm($\gamma, p$)$^{164}$Er.
The reaction rates in the temperature region of 0.1--10 GK are presented in Table~\ref{tab:3}.  Table~\ref{tab:4} shows coefficients of JINA analytical function for these photo-nuclear reaction rates at $T_9=[0.1, 10]$, where $T_9$ is the  temperature in units of 10$^9$ K.
The function is given by:
	\begin{equation}
	N_A\langle \sigma \nu \rangle^\ast=e^{(a_0+a_1T_9^{-1}+a_2T_9^{-1/3}+a_3T_9^{1/3}+a_4T_9+a_5T_9^{5/3}+a_6lnT_9)},
	\end{equation}
where $a_0-a_6$ are the seven coefficients. 
For each reaction, Table~\ref{tab:4} shows reaction rates fitting coefficients calculated using the ‘TALYS\_optimized' model applicable to the $A$$\sim$160 nuclear region (the first row).  The lower and upper bounds corresponding to a 20 \% change in the $R^2$ value (second and third rows).
			
The ratios between the present rates and the JINA recommended rates~\citep{2000ADNDT..75....1R} are shown in Fig.~\ref{fig3}. Uncertainties of the present rates estimated based on the  “Random” are drawn with coloured bands.
For $^{157}$Ho($\gamma, p$)$^{156}$Dy in Fig.~\ref{fig3} (a), the JINA underproduced the rates at $T_9$=[0.4, 0.6] and [2.5, 5]. 
The JINA also underproduced the rates at $T_9$ = [0.3, 6] for $^{159}$Ho($\gamma, p$)$^{158}$Dy. 
For $T=2$--3 GK relevant to the $\gamma$-process, the ratios  for the $^{157}$Ho($\gamma$, $p$)$^{156}$Dy and $^{159}$Ho($\gamma$, $p$)$^{158}$Dy reactions are 0.3 --2.0 and 0.1--0.4, respectively.
Fig.~\ref{fig3} (b) shows that our rates of $^{163}$Tm($\gamma, p$)$^{162}$Er and $^{165}$Tm($\gamma, p$)$^{164}$Er are larger by  factors of 0.5--20.8 and 0.6--25.7 at $T_9=2-3$, respectively.

\section{$\gamma$-process results}\label{SNyields}
			
In this section, we evaluate the relevant changes in the nuclear final abundances of the $\gamma$-process corresponding to the evaluated uncertainties.
A previous study on the carbon-deflagration model of SNe Ia~\citep{Kusakabe2010Production} indicates that several isotopes, such as $^{138}$La, $^{152}$Gd, $^{156,158}$Dy, $^{162,164}$Er, $^{168}$Yb, $^{174}$Hf, $^{180}$Ta, $^{184}$Os and  $^{196}$Hg, could be more produced in a low temperature and density region. Such an enhancement could help to reproduce the solar $p$-nuclei abundances.
Therefore, we set trajectories with peak temperatures $T_{9{\rm p}} =[2,2.9]$ and peak density $\rho_{\rm p} =10^6$ g cm$^{-3}$ for the $\gamma$-process layer in SNe Ia. Such astrophysical condition is also critical for the $\gamma$-process nucleosynthesis at the base of oxygen convective shell or in a C-O shell merger as pointed out in \citet{2001Nucleosynthesis}. We use the analytical profile of temperature and density as $T_9 =T_{9{\rm p}} \exp[-t/(3 \tau)]$ and $\rho =\rho_{\rm p} \exp(-t/\tau)$ with $\tau=1$ s. The initial nuclear composition is Case A1 in ~\cite{Kusakabe2010Production}. The $\gamma$-process is then calculated as in ~\cite{2021ApJ...915...78C} that includes updated rates of $^{16}$O($n$,$\gamma$)$^{17}$O \citep{2020Nuclear} and $^{17}$O($n$,$\gamma$)$^{18}$O \citep{2022ApJ...927...92Z}.
			
\begin{figure}
	\centering
	\includegraphics[width=0.48\textwidth,clip=true,trim=0cm 0cm 0cm 0cm]{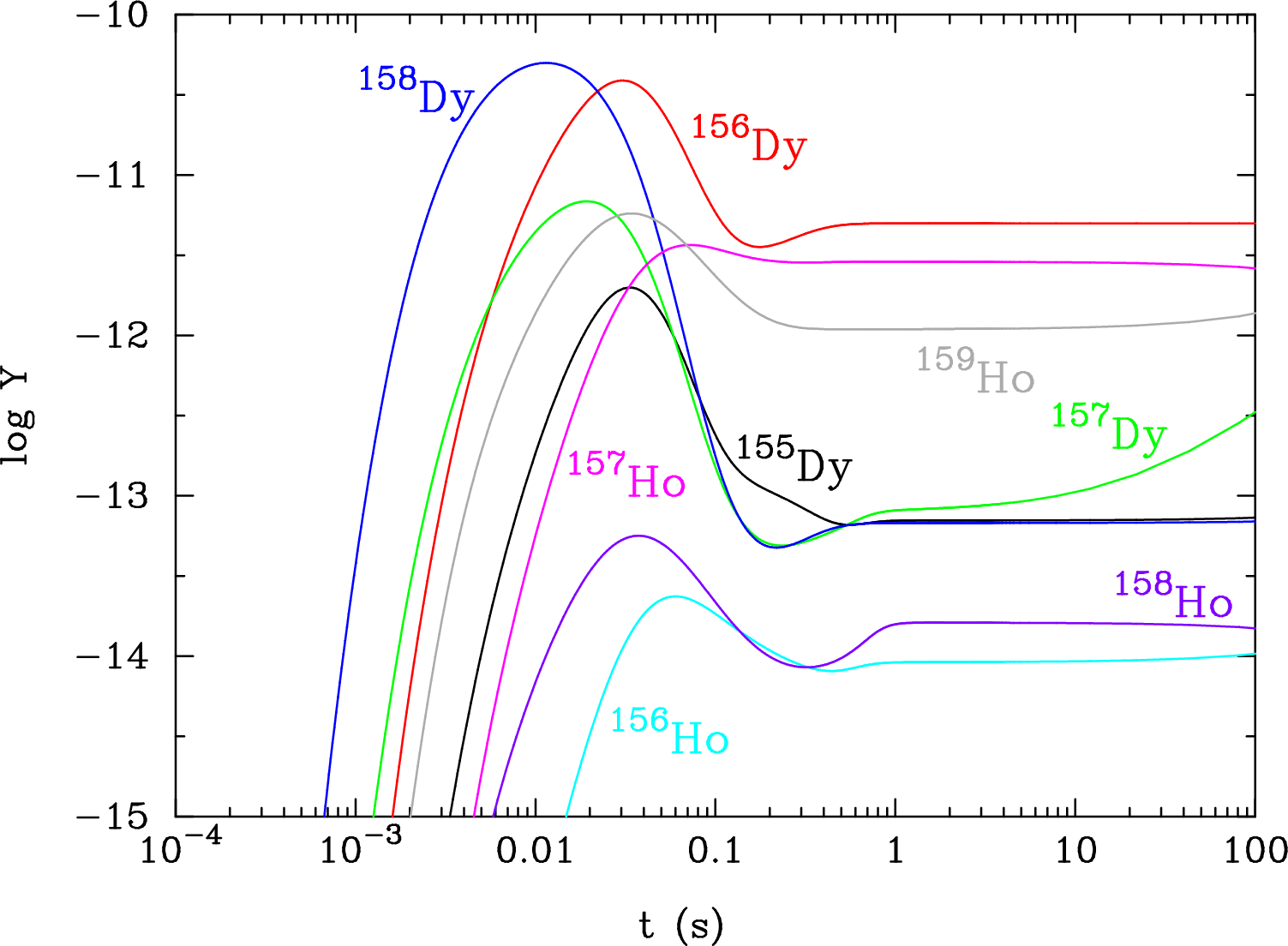}
	\includegraphics[width=0.48\textwidth,clip=true,trim=0cm 0cm 0cm 0cm]{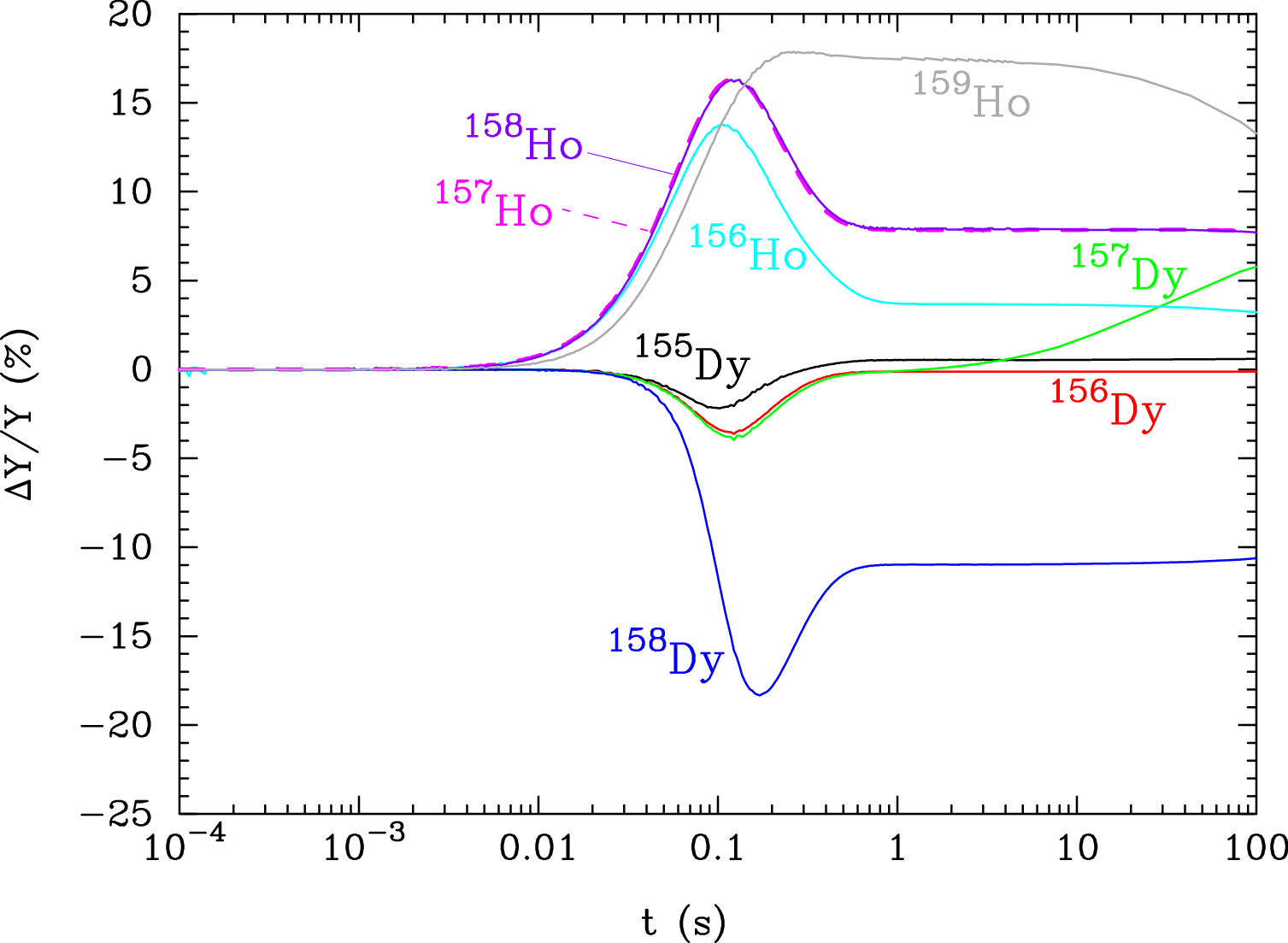}
    \caption{(Color online) Abundances of nuclei around $^{156,158}$Dy (upper panel) in the case with the “optimized” fit rates in this paper and their fractional changes from the case of JINA rates, i.e., $(Y_i -Y_{i\mathrm{JINA}}) /Y_{i\mathrm{JINA}}$ (lower panel) for the profile of $T_{9\mathrm{p}}=2.5$.}
	\label{fig4}
	\end{figure}

Abundances of nuclei around $^{156,158}$Dy in the case with the “optimized” fitted rates of $^{157, 159}$Ho($\gamma$, $p$) for the profile of $T_{9\mathrm{p}}=2.5$ are shown in upper panel of Fig. ~\ref{fig4}. Nuclei are produced at an early time of $t \gtrsim 0.001$ s and then destroyed both via photodisintegration reactions. In addition, their fractional changes from the case of JINA rates, i.e., $(Y_i -Y_{i\mathrm{JINA}}) /Y_{i\mathrm{JINA}}$, are shown in lower panel of Fig.~\ref{fig4}. The adopted rates for the reactions $^{157, 159}$Ho($\gamma$, $p$)$^{156, 158}$Dy are lower than the JINA rates, and the fractional changes in the $^{156, 158}$Dy abundances are negative and those in the $^{157, 159}$Ho abundances are positive.
			
\begin{figure}
	\centering
	\includegraphics[width=0.48\textwidth,clip=true,trim=0cm 0cm 0cm 0cm]{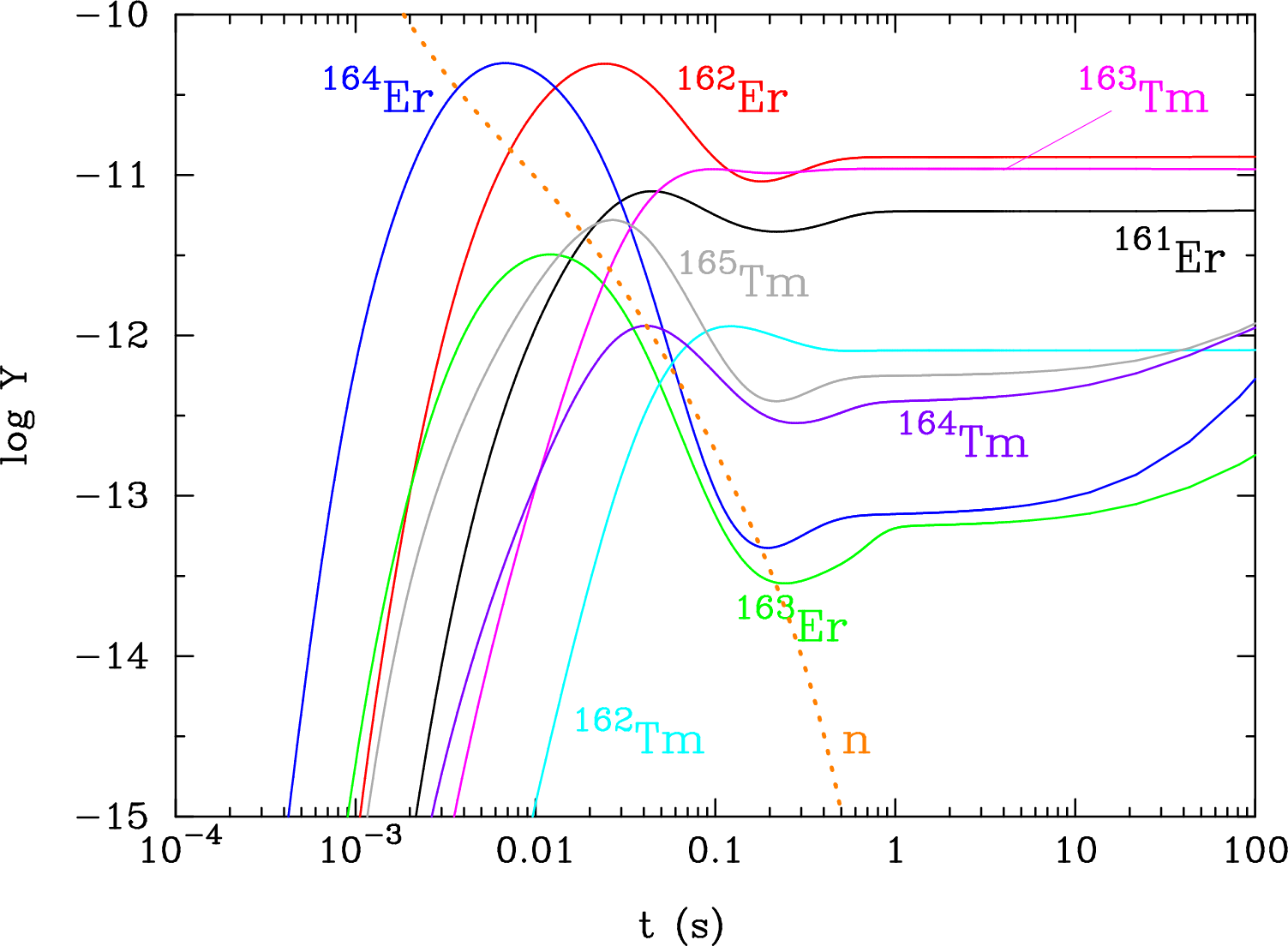}
	\includegraphics[width=0.48\textwidth,clip=true,trim=0cm 0cm 0cm 0cm]{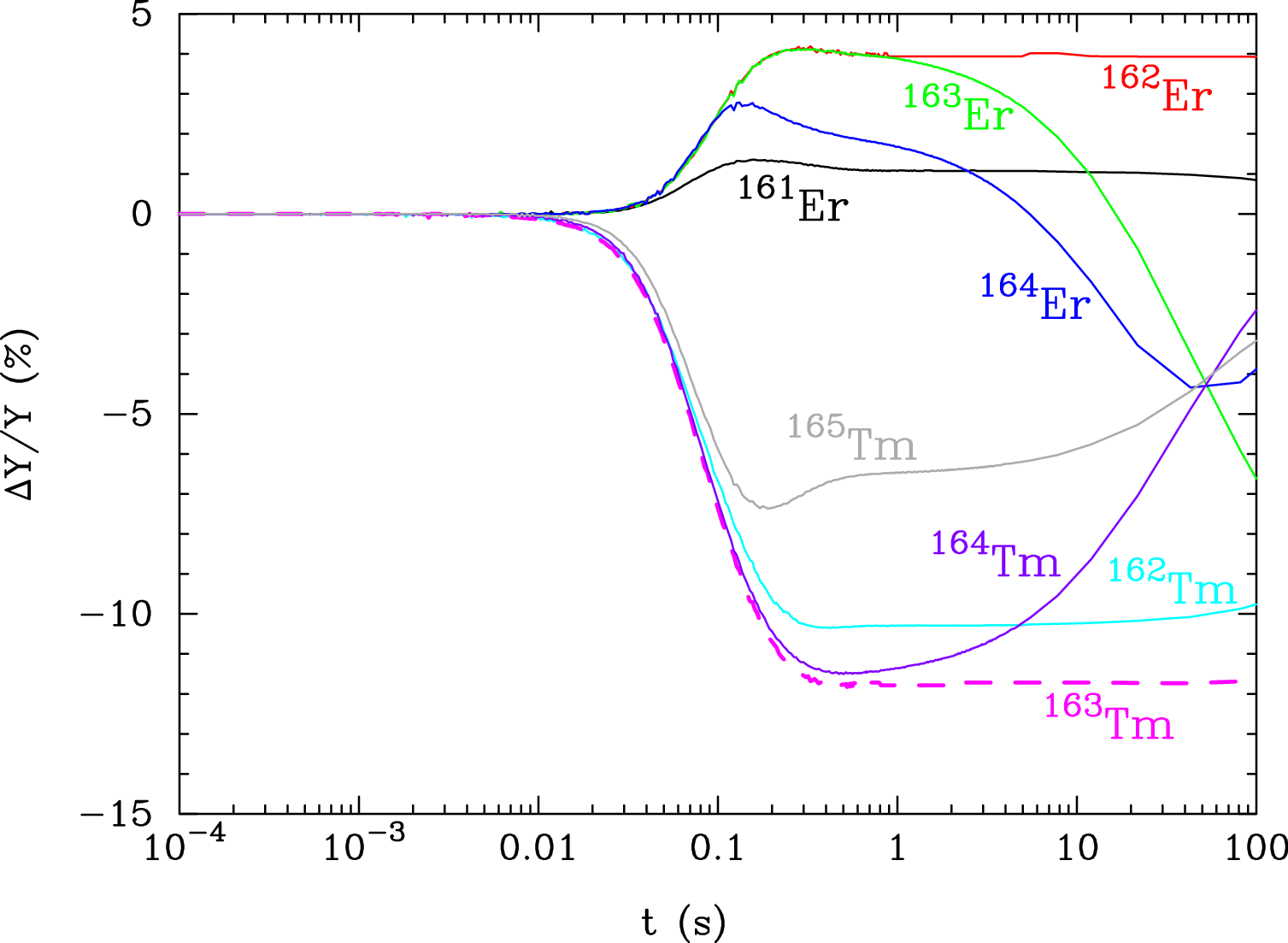}
	\caption{(Color online) The same as in Fig.\ref{fig4} but for nuclei around $^{162,164}$Er.}
	\label{fig5}
	\end{figure}

Upper panel of Fig. \ref{fig5} shows the abundances of nuclei around $^{162,164}$Er in the case with the best fit rates of $^{163,165}$Tm($\gamma$, $p$) for the profile of $T_{9\mathrm{p}}=2.5$. The effect of photodisintegration occurs at an early time. Abundance increases at $t \gtrsim 0.1$ s when the neutron capture becomes more rapid than photodisintegration. In addition, their fractional changes from the case of JINA rates are shown in lower panel of Fig. \ref{fig5}. The adopted rates for the reactions $^{163, 165}$Tm($\gamma$, $p$)$^{162, 164}$Er are higher than the JINA rates. The $^{162}$Er abundance in the best fit case is higher  than those in cases of the JINA rates, and the $^{164}$Er abundance is higher at first but becomes smaller. The $^{163}$Tm abundance is lower, and the $^{165}$Tm abundance is lower but its fractional change increases later.
The $^{164}$Er abundance is higher at first but then becomes smaller. At a later epoch, $^{164}$Er as well as $^{165}$Tm increase.
			
Abundance increases at later time are caused by production via the $\beta^+$-decay and electron capture of nuclei on the proton-rich side. Then, fractional changes tend to approach to zero due to production via the decay and electron capture at a later time when the photodisintegration is not operative.
			
\begin{figure}
	\centering
	\includegraphics[width=0.48\textwidth,clip=true,trim=0cm 0cm 0cm 0cm]{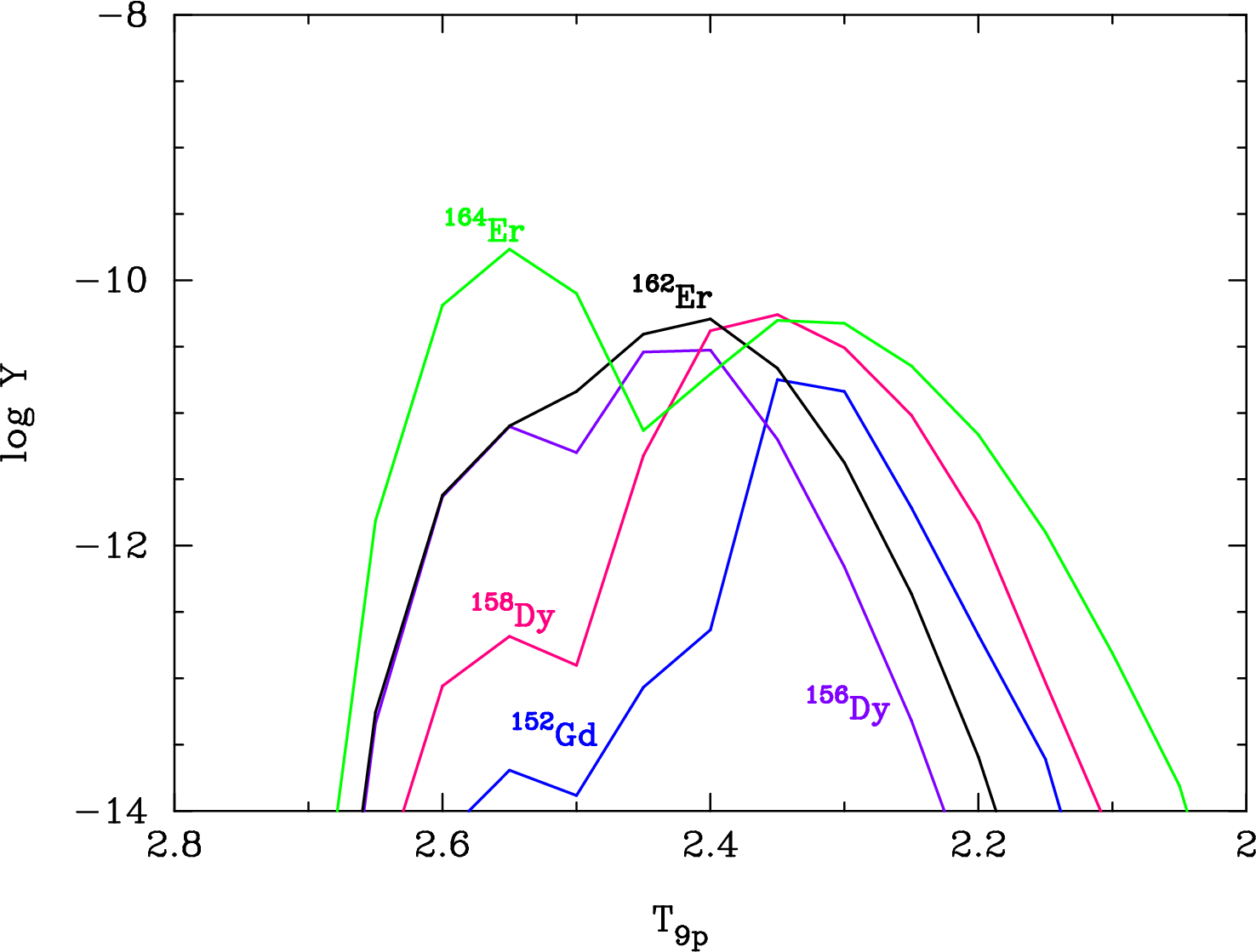}
	\includegraphics[width=0.48\textwidth,clip=true,trim=0cm 0cm 0cm 0cm]{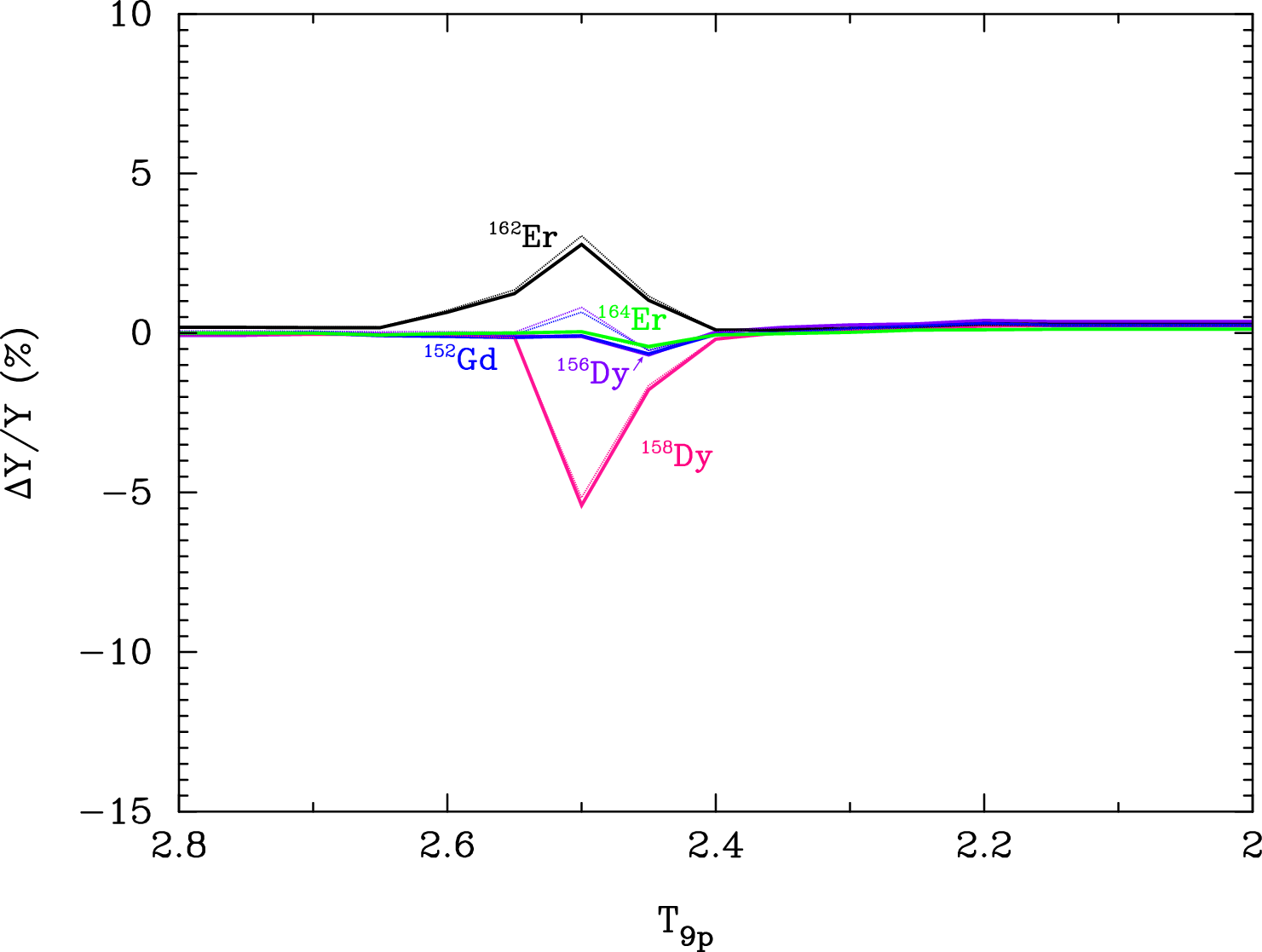}
	\caption{(Color online) Final abundances (upper panel) and their fractional changes from the case of JINA rates (lower panel) in nuclear mole fractions as a function of the peak temperature $T_{9\mathrm{p}}$ for five $p$-nuclei. The values are taken at a longer simulation time, \textit{i.e.}, 1 year after the peak temperature. Thick lines correspond to the best fit rates in this paper, and thin lines correspond to upper and lower limits on the ranges of 20 \% $\chi^2$. }
	\label{fig6}
	\end{figure}

The final abundances and their fractional changes from the case of JINA rates in nuclear mole fractions are shown in upper and lower panels of Fig. \ref{fig6}, respectively, as a function of the peak temperature $T_{9\mathrm{p}}$ for $^{152}$Gd, $^{156,158}$Dy, and $^{162,164}$Er. Thick lines correspond to the best fit rates in this paper, and thin lines correspond to upper and lower limits for two cases.  The upper and lower limits are estimated  from variations in abundances caused by the four ($\gamma$, $p$) reactions for the 20 \% $\chi^2$ case.  Changes in the final abundances between the cases with our best-fit rates and JINA rates are within -5.5--3 \%.
			
\begin{figure}
	\centering
	\includegraphics[width=0.5\textwidth,clip=true,trim=0cm 0cm 0cm 0cm]{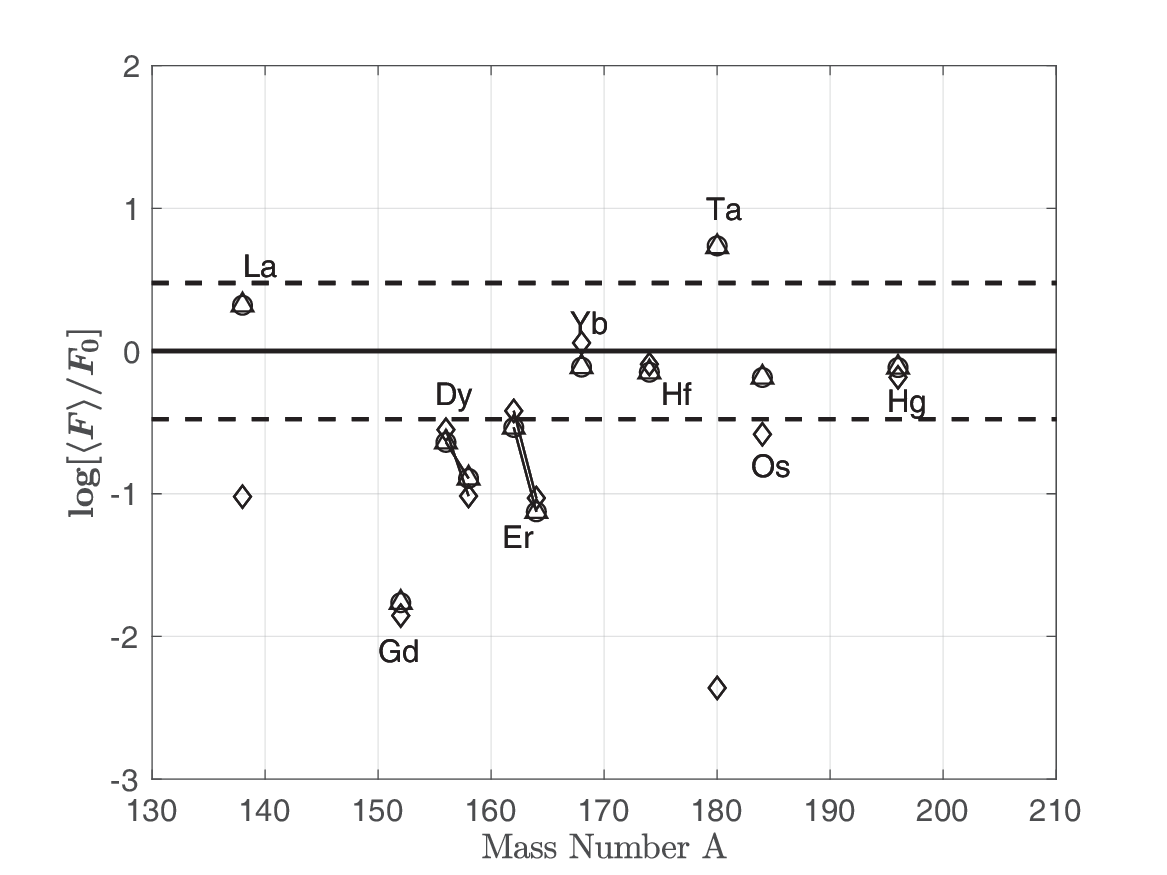}
	\caption{Average overproduction factors calculated by our rates (open circles) compared with JINA (triangles) and~\cite{Kusakabe2010Production} (diamonds). The horizontal solid and dashed lines correspond to $\langle F\rangle /F_0$ equals to 1 and $(3, 1/3)$, respectively.}
	\label{fig7}
	\end{figure}

In Fig.~\ref{fig7} we compare the normalized overproduction factors of $p$-nuclei with $A>130$. The final averaged overproduction factor $\langle F \rangle/F_0$ is calculated for each isotope, where $F_0 = \sum_j \langle F \rangle_j /35$ is the average value for 35 $p$-nuclei. $\langle F \rangle =\sum_i F_i/N$, where $F_i = Y_{i}/Y_\odot$ is the overproduction factor in a trajectory $i$, and $N$ is the total trajectory number. $Y_\odot$ is the solar abundance value taken from \citet{1989GeCoA..53..197A}. \citet{Kusakabe2010Production} calculated more realistic overproduction factors by including the ejection mass of different layers in W7 Model, however, their calculation did not cover the temperature range in this work. Therefore, we added our result to the $\langle F \rangle$ value in their paper to include the contribution from other ejecta of W7 Model.
			
As shown in Fig.~\ref{fig7}, the production of $^{138}$La and $^{180}$Ta are enhanced as expected. This might raise an importance of $\gamma$--process contribution to the solar system abundance of these nuclei whose main production process is claimed to be SNII $\nu$--process ~\citep{1990ApJ...356..272W,2010PhRvC..81e2801H,2014JPhG...41d4007K}, although the calculated results depend on modeling profile of the peak temperature.  Moreover, not all $p$-nuclei with $N>82$ can be enhanced in such a temperature layer, although there are slight increase in $^{152}$Gd, $^{158}$Dy, $^{184}$Os abundances in our calculations (open circles) from JINA (triangles) in Fig.~\ref{fig7}. Also, $^{152}$Gd, $^{156,158}$Dy, $^{162,164}$Er under-production could not be resolved, even from an accurate determination of the $(\gamma, p)$ reaction rates. The circles and triangles overlapped, because the main nucleosynthetic chain under this layer is the $(\gamma,n)$ reactions. Therefore, taking into account the lower temperature layer in the carbon-deflagration model for SNe Ia, $\gamma$-process could not reproduce all the $p$-nuclei abundances. The uncertainty of $(\gamma,p)$ reactions rates only contribute $\le$ 5\% of the final abundances for the nuclei in the mass range $A\sim 160$. Therefore, the underproduction of these nuclei is not explained by a single carbon-deflagration SNe Ia event near the solar system with a near-solar abundance of s-nuclei.
			
\section{Conclusion}\label{conclusion}
For accurate  $\gamma$-process calculations and potential enhancements in theoretical predictions of $p$-nuclei abundances, particularly within the $A$$\sim$160 mass region, we have determined the optimized TALYS physical inputs for this mass range. 
We utilized the available experimental data to constrain the GSF and NLD, which are critical inputs in the HF statistical model.  Additionally, other significant inputs  in TALYS, namely, compound models, pre-equilibrium models, and optical models, were also subjected constraints.
We derive stellar rates using the ‘TALYS\_optimized’ model sets for   $^{157}$Ho($\gamma, p$)$^{156}$Dy, $^{159}$Ho($\gamma, p$)$^{158}$Dy, $^{163}$Tm($\gamma, p$)$^{162}$Dy and $^{165}$Tm($\gamma, p$)$^{164}$Er in the range of $T_9$=[0.1, 10]  for $\gamma$ process calculations. 
Compared to stellar rates provided by JINA database, our results for the best-fit cases of TALYS are different by up to one to two orders of magnitude. 
We applied carbon-deflagration SNe Ia $\gamma$-process for trajectories with a peak temperature $2-2.8$ GK. Our new rates in the four reactions can cause changes in five $p$-nuclei, that is, $^{152}$Gd, $^{156, 158}$Dy,  and $^{162, 164}$Er, with $-5.5\sim3$\% differences in the final abundances at the most sensitive temperature  ($2.4-2.5$ GK). This indicates that if the $(\gamma,p)$ rates for many seed nuclei on the $n$-deficient region are revised substantially, final abundance of $p$-nuclei are possibly to be revised within such a ratio. However, if we averaged the production in the final trajectories, the shifts in final abundances would be $\le$ 5\%. Actually our calculated result showed that the nuclei near mass number $A\sim$ 160 could not be significantly synthesized to reproduce the solar abundance observation. 
Therefore, the solar abundance of those $p$-nuclei might not be explained as a single carbon-deflagration SNe Ia event. 
			
\begin{acknowledgments}
				
We thank two anonymous referees for their valuable suggestions and extend our gratitude to Prof. Yuan Tian for helpful discussions on the reaction rates. This work was supported partially by the National Key R\&D Program of China (Contract No.2022YFA1602401), the National Natural Science Foundation of China (Nos.12305149, 12325506, 11961141004, 12205378, 12335009, 11575018, 12125509, 11961141003), Young Talents Fund of China National Nuclear Corporation under Grant (Nos. FY010260623371, FY212406000901).
				
\end{acknowledgments}

\appendix
\section{Tables}

	\begin{table*}[htbp!]
	\centering
	\caption{Experimental cross sections for $^{160}$Dy($p,\gamma$)$^{161}$Ho and $^{162}$Er($p,\gamma$)$^{163}$Tm compared with TALYS calculations using default and “TALYS\_optimized” inputs . } \label{tab:2}
	\begin{tabular}{ccccc}
		\hline\hline
		&$E_{cm}$     & Exp                & Talys$_\textrm{default}$   & Talys$_\textrm{optimized}$       \\ 
		&(MeV)         & (mb)                 & (mb)              & (mb)                  \\ \hline 
		\multirow{7}{*}{Dy}     &3.34(6)      & 3.25(58)$\times10^{-3}$     &   4.324$\times10^{-3}$    &   3.500$\times10^{-3}$\\ 
		& 4.93(5)     &  0.106(18)        & 0.103                                 & 0.150 \\ 
		&5.43(4)      & 0.152(25)         & 0.155     &    0.197\\ 
		&5.94(4)      &  0.243(40)       & 0.263    &   0.249    \\ 
		&6.44(3)      &  0.310(51)         &0.416     &   0.434 \\ 
		&6.93(3)      &  0.507(84)        & 0.564   &   0.603  \\ \cline{1-5} \hline
		\multirow{9}{*}{Er}      &3.973(3)     & 0.037(5)	       & 0.026     & 0.026\\ 
		&4.471(2)     &  0.173(15)         & 0.132    &  0.133\\ 
		&4.968(2)     & 0.520(44)        & 0.504  &   0.507 \\ 
		&5.465(1)     &  1.918(158)       & 1.562   &    1.572   \\ 
		&5.962(1)     &  4.395(362)     & 2.815   &   3.34 \\ 
		&6.459(1)     & 5.699(471)       & 5.255   &    6.97  \\ 
		&6.956(1))    & 6.601(567)       & 3.133   &   5.66   \\ 
		&7.949(1)     &  8.205(674)       & 2.659  &    5.92 \\ 
		&8.944(1)     &  7.756(652)       & 2.968  &   5.93  \\ \cline{1-5} \hline\hline
	\end{tabular} 
\end{table*}
			
			\begin{table*}[htbp!]
				\footnotesize
				\centering\caption{Selected models in TLAYS varied in this work (first column), corresponding TALYS keywords (second column) and their ranges (third column), and the optimized model parameters reproducing experimental reaction cross section of the $A\approx160$ mass region (fourth column). }\label{tab:1}
				\begin{tabular}{cccc}
					\hline\hline
					Models &  TALYS keywords & Range   & TALYS\_optimized\\
					\hline \rule{0em}{10pt}
					Pre-equilibrium(PE)       &     preeqmode       &   1-4     & 2\\
					Level density models      &     ldmodel         &   1-6     & 2 \\
					Constant Temperature      &     ctmglobal       &   y-n     & n\\
					Mass model                &     massmodel       &   1-3     & /\\
					Width fluctuation         &     widthmode       &   0-3     &2 \\
					Spin cut-off parameter    &     spincutmodel    &   1-2     &2\\
					Shell effects             &     gshell          &   y-n      &y \\
					Excited state in Optical Model &statepot        &   y-n     &n \\
					Spherical Optical Model   &     spherical       &   y-n     &y\\
					Radial matter densities   &     radialmodel     &   1-2     &1\\
					Liquid drop expression    &     shellmodel      &   1-2     &2\\
					Vibrational enhancement   &     kvibmodel       &   1-2     &2\\
					Spin distribution(PE)     &     preeqspin       &   y-n     &n\\
					Kalbach model(pickup)     &     preeqcomplex    &   y-n     &n\\
					Component exciton model   &     twocomponet     &   y-n     &n\\
					Pairing correction(PE)    &     pairmodel       &   1-2     &2\\
					Experimental masses       &     expmass         &   y-n     &y\\
					Gamma-strength function   &     strength        &   1-8     &3\\
					M1 gamma-ray strength function & strengthM1     &   1-2     &1\\
					\hline \hline
				\end{tabular}
			\end{table*}

		\begin{table*}[htbp!]
			\footnotesize
			\centering\caption{Deduced stellar rate (s$^{-1}$) of the $^{157}$Ho($\gamma$, $p$)$^{156}$Dy, $^{159}$Ho($\gamma$, $p$)$^{158}$Dy, $^{163}$Tm($\gamma$, $p$)$^{162}$Er and $^{165}$Tm($\gamma$, $p$)$^{164}$Er in the range of $T_9$ = [0.1,10]. }\label{tab:3}
			\begin{tabular}{ccccc}
				\hline\hline
				$T_9$ &  $^{157}$Ho($\gamma$, $p$)$^{156}$Dy & $^{159}$Ho($\gamma$, $p$)$^{158}$Dy &  $^{163}$Tm($\gamma$, $p$)$^{162}$Er &  $^{165}$Tm($\gamma$, $p$)$^{164}$Er\\
				\hline \rule{0em}{10pt}
				0.10  & 1.67$\times10^{-217}$ &   1.36$\times10^{-248}$ &  1.52$\times10^{-222}$ &  3.07$\times10^{-252}$  \\
				0.15 & 6.20$\times10^{-147}$ &   1.34$\times10^{-167}$ &  1.96$\times10^{-150}$ &  2.58$\times10^{-170}$  \\
				0.20    & 1.29$\times10^{-111}$ &   4.68$\times10^{-127}$ &  2.38$\times10^{-114}$ &  2.71$\times10^{-129}$  \\
				0.30     & 1.87$\times10^{-75}$ &   1.26$\times10^{-85}$ &  2.38$\times10^{-77}$ &  3.28$\times10^{-87}$  \\
				0.40  & 2.13$\times10^{-56}$ &   6.59$\times10^{-64}$ &  7.23$\times10^{-58}$ &  3.80$\times10^{-65}$  \\
				0.50 & 1.93$\times10^{-44}$ &   2.41$\times10^{-50}$ &  1.20$\times10^{-45}$ &  2.14$\times10^{-51}$  \\
				0.60  & 3.71$\times10^{-36}$ &   5.43$\times10^{-41}$ &  3.49$\times10^{-37}$ &  6.25$\times10^{-42}$  \\
				0.70  & 4.91$\times10^{-30 }$&   4.12$\times10^{-34}$ &  6.24$\times10^{-31}$ &  5.65$\times10^{-35}$  \\
				0.80  & 2.68$\times10^{-25 }$&   8.22$\times10^{-29}$ &  4.26$\times10^{-26}$ &  1.28$\times10^{-29}$  \\
				0.90   & 1.65$\times10^{-21 }$&   1.38$\times10^{-24}$ &  3.13$\times10^{-22}$ &  2.38$\times10^{-25}$  \\
				1.00 & 2.12$\times10^{-18 }$&   3.91$\times10^{-21}$ &  4.65$\times10^{-19}$ &  7.37$\times10^{-22}$  \\
				1.50 & 1.62$\times10^{-8 }$&   2.97$\times10^{-10}$ &  5.39$\times10^{-9}$ &  7.56$\times10^{-11}$  \\
				2.00 & 4.03$\times10^{-3 }$&   2.15$\times10^{-4}$ &  1.64$\times10^{-3}$ &  6.77$\times10^{-05}$  \\
				2.50 & 1.19$\times10^{1 }$&   1.12$\times10^{0}$ &   5.34$\times10^{0}$&  4.12$\times10^{-1}$  \\
				3.00 & 3.25$\times10^{3 }$&  4.09$\times10^{2 }$  &  1.52$\times10^{3}$ & 1.67$\times10^{2}$  \\
				3.50 & 2.07$\times10^{5 }$&  2.90$\times10^{4}$ &  9.62$\times10^{4}$ &  1.28$\times10^{4}$  \\
				4.00 & 4.92$\times10^{6}$&  6.94$\times10^{5}$ &  2.20 $\times10^{6}$ &  3.23$\times10^{5}$  \\
				5.00 & 4.07$\times10^{8}$&  5.17$\times10^{7}$ &  1.62$\times10^{8}$ &  2.64$\times10^{7}$  \\
				6.00 & 6.95$\times10^{9}$&  8.13$\times10^{8}$ &  2.50$\times10^{9}$ &  4.58$\times10^{8}$  \\
				7.00 & 4.88$\times10^{10}$&  5.94$\times10^{9}$ &  1.69$\times10^{10}$ &  3.70$\times10^{9}$  \\
				8.00 & 2.03$\times10^{11}$&  2.82$\times10^{10}$ &  7.28$\times10^{10}$ &  1.90$\times10^{10}$  \\
				9.00 & 5.90$\times10^{11}$&  9.45$\times10^{10}$ &  2.24$\times10^{11}$ &  6.53$\times10^{10}$  \\
				10.0 & 1.28$\times10^{12}$&  2.27$\times10^{11}$ & 5.04$\times10^{11}$ &  1.56$\times10^{11}$  \\
				\hline \hline
			\end{tabular}
		\end{table*}

		\begin{table*}[htbp!]
			\centering
			\caption{Coefficients of JINA analytical function for photo-nuclear 
				reaction rates of  $^{157}$Ho($\gamma, p$)$^{156}$Dy, $^{159}$Ho($\gamma, p$)$^{158}$Dy,  $^{163}$Tm($\gamma, p$)$^{162}$Er and $^{165}$Tm($\gamma, p$)$^{164}$Er in the range of $T_9$=[0.1, 10] } \label{tab:4}
			\begin{tabular}{cccccccccc}
				\hline\hline
				Coefficient	& & &$a_0$     & $a_1$     & $a_2$   &$a_3$ 	&$a_4$       &$a_5$      & $a_6$                  \\ \hline 
				\multirow{3}{*}{$^{157}$Ho($\gamma, p$)$^{156}$Dy } &	&optimized	&306.8256	&80.36393	&-2873.602	&2549.025	&-108.4133	&5.107397	&-1562.009 \\ \cline{2-10} 
				&\multirow{2}{*}{20\% $\chi^2$}  &lower limit  	&305.9222	&83.26270	&-2947.910	&2624.749	&-112.0218	&5.301037	&-1605.827\\
				&&upper limit	&391.2658	&73.92737	&-2639.568	&2216.876	&-86.70362	&3.531499	&-1398.746 \\  \hline
				\multirow{3}{*}{$^{159}$Ho($\gamma, p$)$^{158}$Dy } &	&optimized	&118.0938	&80.72147	&-3286.785	&3181.858	&-148.6443	&7.765966 &-1867.468\\ \cline{2-10} 
				&\multirow{2}{*}{20\% $\chi^2$}  &lower limit	 &82.09172	&74.91393	&-3191.848	&3127.722	&-147.6481	&7.777003	&-1824.572\\
				&&upper limit	 &243.9814	&84.48611	&-3273.240	&3024.010	&-132.5265	&6.339217	&-1820.269\\ \hline	
				\multirow{3}{*}{$^{163}$Tm($\gamma, p$)$^{162}$Er} &	&optimized	&50.44470	&-3.327696	&-1099.667	&1058.359	&-50.53199	&2.507785	&-609.8706\\ \cline{2-10}				
				&\multirow{2}{*}{20\% $\chi^2$}  &lower limit	 &45.66944	&-5.379221	&-1053.155	&1017.024	&-48.79684	&2.418792	&-584.2264\\				
				&&upper limit	 &134.6496	&2.013127	&-1184.373	&1046.686&-42.71171	&1.559894	&-635.1377\\ \hline
				\multirow{3}{*}{$^{165}$Tm($\gamma, p$)$^{164}$Er} &	&optimized	&474.1743	&-73.44261
				&935.1337	&-1468.801	&90.64918	&-6.253037	&721.2172\\ \cline{2-10}								
				&\multirow{2}{*}{20\% $\chi^2$}  &lower limit	 &471.0940	&-73.37822	&931.5245	&-1461.644 &90.06996	&-6.207963	&718.2022\\
				&&upper limit	 &524.4451	&-74.26482	&985.1584	&-1576.997	&100.2338	&-7.085249	&764.8725\\
				\hline\hline
			\end{tabular} 
		\end{table*}l

	\end{document}